\documentclass%
[aps,prl,superscriptaddress,twocolumn,showpacs,preprintnumbers,floatfix]{revtex4}
\usepackage{hyperref}
\usepackage{amsmath}
\usepackage{amsfonts}
\usepackage{amssymb}
\usepackage{comment}
\usepackage{graphicx}
\begin{document}
\title{Theory of ground state cooling of a mechanical
  oscillator using dynamical back-action} 
\author{I.~Wilson-Rae}
\affiliation{Technische Universit\"{a}t M\"{u}nchen, D-85748 Garching,
  Germany.} 
\author{N.~Nooshi}
\affiliation{Max Planck\ Institut f\"{u}r Quantenoptik, D-85748
  Garching,\ Germany.}
\author{W.~Zwerger}
\email[]{zwerger@ph.tum.de}
\affiliation{Technische Universit\"{a}t M\"{u}nchen, D-85748 Garching,
  Germany.} 
\author{T.J.~Kippenberg}
\email[]{tjk@mpq.mpg.de}
\affiliation{Max Planck\ Institut f\"{u}r Quantenoptik, D-85748
  Garching,\ Germany.} 
\date{\today}

\begin{abstract}
 A quantum theory of cooling of a mechanical oscillator by radiation
 pressure-induced dynamical back-action is developed, which is
 analogous to sideband cooling of trapped ions. We find that final
 occupancies well below unity can be attained when the mechanical
 oscillation frequency is larger than the cavity linewidth. It is
 shown that the final average occupancy can be retrieved directly from the
 optical output spectrum.

\end{abstract}
\maketitle

\begin{figure}[ptb]
  \centering\includegraphics[width=2.8in]{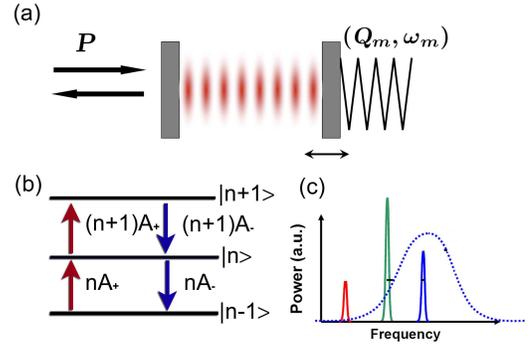}\caption{(color
    online) (a) Fabry-P\'{e}rot equivalent of a mechanical eigenmode
    (frequency $\omega_m/2\pi$ and quality factor $Q_m$) coupled to an
    optical cavity mode. (b) Processes that decrease or increase the
    mechanical eigenmode's quantum number $n$. Cooling occurs by
    scattering into anti-Stokes modes, whereas heating (amplification)
    proceeds via Stokes scattering. (c) Illustration that compares the
    output power spectral density (solid lines), exhibiting motional
    sidebands (blue and red), with the cavity absorption in the
    absence of optomechanical coupling (dashed).\label{fig:FP}}%
\end{figure}

Mesoscopic mechanical oscillators are currently attracting interest
due to their potential to enhance the sensitivity of displacement
measurements \cite{Abramovici1992} and to probe the quantum to
classical transition of a macroscopic degree of freedom
\cite{Mancini2002,Schwab2005}. A prerequisite for these applications
is the capability of initializing an oscillator with a long phonon
lifetime in its quantum ground state. So far this has not been
demonstrated because the combination of sufficiently high mechanical
frequencies ($\omega_m/2\pi$) and quality factors in the relevant
regime $\hbar\omega_m\gg k_B T$ has not been reached
\cite{Schwab2005}.  In contrast, in atomic physics laser cooling has
enabled the preparation of motional ground states
\cite{Wineland1979,Stenholm1986Leibfried2003}. This has prompted
researchers to study means of cooling a single mechanical resonator
mode directly using laser radiation. Early work demonstrated cooling
of a mechanical degree of freedom of a Fabry-P\'{e}rot mirror using a
radiation pressure force controlled by an electronic feedback scheme
\cite{Mancini1998,Cohadon1999}, in analogy to stochastic cooling. In
contrast, the radiation pressure induced coupling of an optical cavity
mode to a mechanical oscillator [cf.~Fig.~\ref{fig:FP}(a)] can give
rise to self-cooling via \textit{dynamical back-action}
\cite{Braginsky19922001}. In essence, the cavity delay induces
correlations between the radiation pressure force and the thermal
Brownian motion that lead to cooling or amplification, depending on
the laser detuning. In a series of recent experiments, these effects
have been used to cool a single mechanical mode
\cite{Gigan2006,Pinard2006,Schliesser2006}. While classical and
semiclassical analysis of dynamical back-action have been developed
\cite{Braginsky2002Marquardt2006,Kippenberg2005}, the question as to
whether \emph{ground state cooling} is possible has not been
addressed.

Here a quantum theory of cooling via dynamical back-action is
presented. We find that final occupancies below unity can indeed be
attained when the optical cavity's lifetime is comparable to or
exceeds the mechanical oscillation period. Along these lines, an
analogy between this mechanism and the sideband cooling of
trapped ions in the Lamb-Dicke regime is elucidated
\cite{Stenholm1986Leibfried2003}. In our setting the optical cavity mode
plays the role of the ion's pseudospin mediating the frequency
up-conversion underlying the cooling cycle.  Finally, we discuss how
the average phonon occupancy can be retrieved from the spectrum of the
optical cavity output. We note that these results can be applied to a
wide range of experimental realizations of cavity self-cooling
\cite{Gigan2006,Schliesser2006,Kleckner2006}.

We treat the laser driven optical cavity mode coupled to the
mechanical resonator mode as an open quantum system and adopt a
rotating frame at the laser frequency $\omega_L$. The system
Hamiltonian is given by \cite{Law1995,Walls1994}
\begin{align}
\label{eq:Hprime}H^{\prime}=  &  -\hbar\Delta_L^{\prime}a_{p}^{\dag}%
a_{p}^{\vphantom\dag} + \hbar\eta\omega_{m} a_{p}^{\dag}a_{p}^{\vphantom\dag
}\left(a_{m}^{\vphantom\dag}+a_{m}^{\dag}\right) \nonumber\\
&  + \hbar\textstyle{\frac{\Omega}{2}}\!\left(  a_{p}^{\vphantom\dag}%
+a_{p}^{\dag}\right)  + \hbar\omega_{m} a_{m}^{\dag}a_{m}^{\vphantom\dag}\,.
\end{align}
Here $a_{p}$ ($a_{m}$) is the annihilation operator for the optical
(mechanical) oscillator, $\omega_{p}$ ($\omega_{m}$) its angular
frequency and $\Delta_L^{\prime}$ the laser detuning from the optical
resonance.  We have also introduced the driving amplitude
$\Omega\equiv2\sqrt {P/\hbar\omega_L\tau_\mathrm{ex}}$, where $P$ is
the input laser power and $1/\tau _\mathrm{ex}$ the photon decay rate into
the associated outgoing modes (\emph{e.g.}  optical fiber modes
\cite{Spillane2003}). The optomechanical coupling via radiation
pressure can be characterized by the dimensionless parameter
$\eta\equiv(\omega_{p}/\omega_{m})(l_m/L)$; with
$l_m=\sqrt{\hbar/2m\omega_{m}}$ the zero point motion of the
mechanical resonator mode, $m$ its effective mass, and $L$ an
\emph{effective} length that depends on how the radiation pressure
force affects the optical cavity. For typical materials and dimensions
\cite{Schliesser2006} one obtains $\eta\sim 10^{-4}$.

The optical cavity losses and the intrinsic dissipation of the
mechanical resonator are characterized, respectively, by the cavity
lifetime $\tau$ and the mechanical quality factor $Q_m$. These give
rise to a dissipative contribution to the Liouvillian
$\mathcal{L^{\prime}}_D$ (i.e.~$\mathcal{L^{\prime}}=
-\frac{i}{\hbar}[ H^{\prime},\ldots] + \mathcal{L^{\prime}}_D$) that
is of Lindblad form \footnote{This requires
  rotating-wave-approximations that are warranted in the parameter
  regime of interest.} with collapse operators \cite{Gardiner2004,Walls1994}:
$\sqrt{1/\tau}a_{p}$,
$\sqrt{\gamma_{m}n(\omega_{m})}a_{m}^{\dagger}$ and
$\sqrt{\gamma_{m}[n(\omega_{m})+1]}a_{m}^{\vphantom\dagger}$.
\begin{comment}
\begin{align}\label{eq:Lprime}
\mathcal{L^{\prime}}_D\rho=  &  \textstyle{\frac{1}{2\tau}}\!\left(  2
a_{p}^{\vphantom\dag}\rho a_{p}^{\dag}-
a_{p}^{\dag}a_{p}^{\vphantom\dag}\rho- \rho
a_{p}^{\dag}a_{p}^{\vphantom\dag}\right) \nonumber\\ & + \omega_{m}
\textstyle{\frac{n\left( \omega_{m}\right) }{2Q_{m}} }\!\left( 2
a_{m}^{\dagger}\rho a_{m}^{\vphantom\dagger} - a_{m}
^{\vphantom\dagger}a_{m}^{\dagger}\rho- \rho a_{m}^{\vphantom\dagger}
a_{m}^{\dagger} \right) \\ & +\omega_{m}\textstyle{\frac{n\left(
\omega_{m}\right) +1}{2Q_{m}}}\!\left( 2
a_{m}^{\vphantom\dagger}\rho a_{m}^{\dagger} - a_{m}^{\dagger
}a_{m}^{\vphantom\dagger}\rho- \rho
a_{m}^{\dagger}a_{m}^{\vphantom\dagger }\right) \,,\nonumber
\end{align}
\end{comment}
Here $\gamma_m=\omega_{m}/Q_{m}$ is the mechanical oscillator's
natural linewidth and $n(\omega_{m})$ its Bose number at the
environmental temperature. We will focus on the regime
$[n(\omega_{m})+1]\gamma_{m}\ll \omega_{m}$, $\gamma_{m}\ll1/\tau$ and
$\eta,\eta|\alpha|\ll1,\,1/\omega_{m}\tau$. The first condition is
necessary for ground state cooling (see below), the second one is
satisfied in all recent experiments
\cite{Gigan2006,Schliesser2006,Pinard2006}, and the last one, given
the smallness of $\eta$, will hinge on having a sufficiently low input
power.

%, as will be borne out below,

To study the dynamics generated by the Liouvillian
$\mathcal{L^{\prime}}$ it proves useful to apply a shift to the modes'
normal coordinates: $a_{p}\rightarrow a_{p}+\alpha$, $a_{m}\rightarrow
a_{m}+\beta$ with the c-numbers $\alpha$ and $\beta$ chosen to cancel
out all the linear terms in the transformed Liouvillian
$\mathcal{L^{\prime} }\rightarrow\mathcal{L}$. To zeroth order in the
small parameters $\eta$ and $1/Q_{m}$ we have \footnote{The
corrections do not affect the lowest order contributions in
$\eta^{2},\,1/Q_{m}$ to the cooling and heating rates.}:
$\alpha\approx\Omega\tau/(2\tau\Delta_L^{\prime}+i)$, $\beta\approx
-\eta|\alpha|^{2}$. We include the radiation pressure
induced optical resonance shift into the effective detuning
$\Delta_L^{\prime}+2\eta^{2}|\alpha|^{2}\omega_{m}%
\rightarrow\Delta_L$ and perform the additional canonical
transformation $a_{p}\rightarrow (\alpha/|\alpha|)a_{p}$. While the
dissipative part of the Liouvillian $\mathcal{L^{\prime}}$ remains
invariant, the Hamiltonian transforms into
\begin{align}
H= & -\hbar\Delta_L a_{p}^{\dag}a_{p}^{\vphantom\dag}+\hbar\omega
_{m}a_{m}^{\dag}a_{m}^{\vphantom\dag}+\hbar\eta\omega_{m}\left[
a_{p}^{\dag }a_{p}^{\vphantom\dag}\right.  \nonumber\label{eq:H}\\ &
\left.  +\,|\alpha|\!\left( a_{p}^{\vphantom\dag}+a_{p}^{\dag}\right)
\right] \left( a_{m}^{\vphantom\dag}+a_{m}^{\dag}\right) \,.
\end{align}

Henceforth we will refer to the primed representation
(\ref{eq:Hprime}) as the \textquotedblleft physical\textquotedblright
one and to the unprimed representation (\ref{eq:H}) as the
\textquotedblleft shifted\textquotedblright one. The smallness of
$\eta^{2}$ and $[n(\omega _{m})+1]/Q_{m}$ imply a wide separation
between the timescales for cooling and heating the mechanical
oscillator and those characterizing the dynamics of the optical cavity
mode and the mechanical oscillation period. Thus, the electromagnetic
environment (including the optical cavity) can be regarded as a
structured reservoir with which the mechanical mode interacts
perturbatively [cf.~Fig.~\ref{fig:FP}(b),(c)]. This prompts us to
derive a \textquotedblleft generalized quantum
optical\textquotedblright\ master equation for the reduced density
matrix \cite{Gardiner2004} of the latter:
$\mu=\textrm{Tr}_{p}\{\rho\}$. In our context, such a derivation can
also be viewed as an adiabatic elimination
\cite{Gardiner2004} of the optical cavity in the presence of fast
rotating terms ($\propto e^{\pm i\omega_{m}t}$) in the optomechanical
interaction. We note that while in the physical representation the
steady state average occupancy of the optical cavity is given by
$|\alpha|^2$, in the shifted one its steady state is simply the vacuum
$|0\rangle_{p}$.  Thus, we obtain
\begin{align}
  \dot{\mu}= & -i\left[\omega_{m}
    a_{m}^{\dagger}a_{m}^{\vphantom\dagger},\mu\right]
    +\textstyle{\frac{1}{2}}\!\left\{\gamma_{m}\left[n\left(
    \omega_{m}\right)\,+\,1\right] + A_{-}\right\}
\nonumber\label{eq:master}\\ &
    \times\left(2a_{m}^{\vphantom\dagger}\mu
    a_{m}^{\dagger}-a_{m}^{\dagger}a_{m}^{\vphantom\dagger}\mu-\mu
    a_{m}^{\dagger }a_{m}^{\vphantom\dagger}\right) +
    \textstyle{\frac{1}{2}}\!\left[ \gamma_{m}n\left(
    \omega_{m}\right) \right.\nonumber\\ & \left. +\,
    A_{+}\right]\left(2a_{m}^{\dagger}\mu
    a_{m}^{\vphantom\dagger}-a_{m}
    ^{\vphantom\dagger}a_{m}^{\dagger}\mu-\mu a_{m}^{\vphantom\dagger}
    a_{m}^{\dagger}\right)\,.
\end{align}
In the first term we have redefined $\omega_m$ to include the
light-induced shift of the mechanical frequency. The second and third
terms correspond, respectively, to cooling and heating induced by the
coupling to the thermal bath (contributions $\propto \gamma_m$) and by
inelastic laser light scattering processes [cf.~Fig.~\ref{fig:FP}(b)]
with rates
\begin{equation}
A_{\mp}=\eta^{2}\frac{4\Omega^{2}}{4\tau^{2}\Delta_L^{2}+1}\,\frac
{\omega_{m}^{2}\tau^{3}}{4\tau^{2}\left( \Delta_L\pm\omega_{m}\right)
^{2}+1}\,.\label{eq:A}
\end{equation}

%which are proportional to the input power $P$.
% which is given by the Hilbert transform of $\Gamma(\omega_m)/2$.

%(dispersive contribution)
%\footnote{The
%detuning-independent prefactor can be related to the finesse $F$ of an
%equivalent Fabry-P\'erot (cf.~Fig.~\ref{fig:FP}.a) --- defined by an
%optical cavity length $L$ --- via the relation:
%$\eta^2\Omega^2\omega_m^2\tau^3\approx2\omega_p n^2 F^2\tau P/\pi^2 m
%\omega_m c^2\tau_\textrm{ex}$ .}

In the shifted representation it is simple to understand these cooling
and heating rates in terms of perturbation theory in the small
parameters $\eta|\alpha|$ and $\eta$ [cf.~Eq.~(\ref{eq:H})]. To
lowest order in $\eta$ only the states $|0\rangle_{p}$ and
$|1\rangle_{p}$ participate yielding the same results as for an
equivalent dissipative two level system. Denoting by $|n\rangle$ the
number states of the mechanical oscillator we have anti-Stokes
(Stokes) processes in which the transition
$|0\rangle_{p}|n\rangle\rightarrow|1\rangle _{p}|n-1\rangle$
($|0\rangle_{p}|n\rangle\rightarrow|1\rangle_{p}|n+1\rangle$) followed
by the decay of the cavity photon leads to cooling
(amplification). This scenario is thus similar to the laser cooling of
a trapped ion in the Lamb-Dicke regime
\cite{Neuhauser1978,Wineland1979,Stenholm1986Leibfried2003}, or of a
nanomechanical resonator coupled to an ``artificial atom''
\cite{Wilson2004Martin2004} or an ion \cite{Zoller2004}. An important
caveat in this analogy is that there is no external driving for
$\eta=0$. Furthermore, though the parameter $\eta^{2}$ will play a
role reminiscent of the Lamb-Dicke parameter --- determining for
example the relative weights of the sidebands --- the efficiency of
the cooling process will depend solely on $\eta^{2}|\alpha|^2$ and
Eqs.~(\ref{eq:A}) will remain valid for arbitrary $\Omega$ provided
$\eta^{2}$ is sufficiently small. This absence of a ``direct'' driving
amplitude also implies that the cubic term in Hamiltonian (\ref{eq:H})
does not contribute to the master equation (\ref{eq:master}) as it
only generates terms that are higher order in $\eta^{2}$. Thus to
lowest order there is no \textquotedblleft diffusive
channel\textquotedblright and the theory is equivalent to a quadratic
Liouvillian \footnote{While Eq.~(\ref{eq:A}) is valid for arbitrarily
low laser intensities it coincides with its counterpart in a classical
treatment of the canonical variables as used in
Ref.~\cite{Schliesser2006}.}.

Henceforth we focus on the regime $\Delta_L<0$ for which there is a
net laser cooling rate $\Gamma=A_{-}-A_{+}>0$. In this regime
Eq.~(\ref{eq:master}) has a well defined steady state that transformed
back to the physical representation yields a shifted thermal state.
The corresponding steady state average occupancy, to which the system
converges on the timescale $1/(\Gamma+\gamma_{m})$, is given by
$\langle a_{m}^{\dagger}a_{m}^{\vphantom\dagger}\rangle_\mathrm{SS}
=n_{f}+|\beta|^{2}$ with $n_{f}=[\gamma_{m}n(\omega_{m})+
A_{+}]/(\gamma_{m}+\Gamma)$ and
$|\beta|^{2}\!=\!\eta^{2}\Omega^{4}\tau^{4}/(4\tau^{2}\Delta_L^{2}+1)^{2}%
$. However the final temperature should be defined in terms of $n_{f}$
as the other contribution arises from a coherent shift. In fact, under
the conditions underpinning our approximations, the latter could be
undone by a ``slow'' switch-off of the laser.

%suitable adiabatic

\begin{figure}[t]
  \centering\includegraphics[width=2.8in]{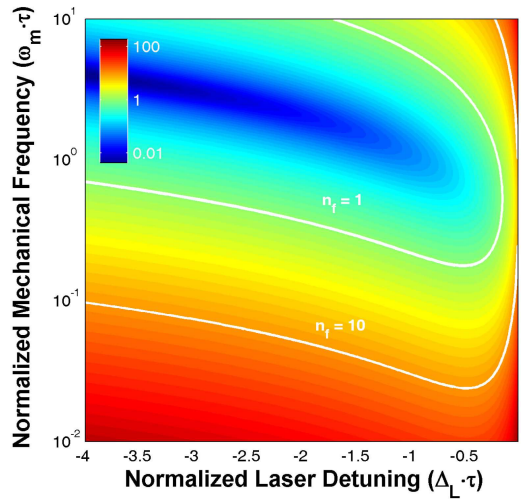}
  \caption{(color online) Final (steady state) average phonon number
    $\tilde{n}_f$ as a function of normalized laser detuning
    ($\Delta_L\tau$) and normalized mechanical angular frequency
    ($\omega_{m}\tau$). The contour-lines indicate the values
    $\tilde{n}_f = 1$ and $10$, respectively. Ground state cooling is
    only possible in a finite detuning window and for
    $\omega_{m}\tau>1/\sqrt{32}$. \label{fig:FT}}
\end{figure}

As we start from thermal equilibrium, initially the number of phonons
is given by $n_{i}=n(\omega_{m})$. Thus, from the expression for
$n_{f}$ it is clear that for appreciable cooling
(\emph{i.e.}~$n_{f}\ll n_{i} $) we need $\Gamma\gg\gamma_{m}$.
In this regime 
\begin{equation}\label{eq:nf}
n_{f}=\left[\frac{\gamma_{m}}{\Gamma}\,n_i +
\tilde{n}_f\right]\left[1 +
\mathcal{O}\left(\frac{\gamma_{m}}{\Gamma}\right)\right]\,,
\end{equation}
with
\begin{equation}\label{eq:ntilde}
\tilde{n}_f\equiv\frac{A_+}{\Gamma}=\frac{4\tau^{2}
\left(\Delta_L+\omega_{m}\right)^{2}
+1}{16\tau^{2}\omega_{m}\left(-\Delta_L\right) }.
\end{equation}
Note that from Eq.~(\ref{eq:A}) it follows that $\Gamma$ is exactly
proportional to the input power $P$. Since there are two terms in
Eq.~(\ref{eq:nf}), there are two regimes depending on which
contribution dominates the behavior of $n_{f}$.  In the first regime
heating is dominated by the intrinsic dissipation of the mechanical
resonance and the final average occupancy is proportional to the
initial one. 
This behavior has already been demonstrated experimentally
\cite{Pinard2006,Schliesser2006,Kleckner2006}. On the other hand for
sufficiently low $\gamma_{m}$ and input laser power, the heating is
dominated by the scattering of laser light. In this regime the final
occupancy is given by $\tilde{n}_f$ (cf.~Fig.~\ref{fig:FT}). Thus the
optimal value of $n_{f}$ solely depends on the product
$\omega_{m}\tau$ and is found by minimizing with respect to the
normalized detuning $\delta\equiv \tau\Delta_L$ (\emph{i.e.}~it is
power independent).  This yields
$\delta_\textrm{opt}=-\sqrt{1+4\omega_{m}^{2}\tau^{2}}/2$, implying the
\textit{fundamental temperature limit}:
\begin{equation}
  \tilde{n}_\textrm{TL}=\min\left\{\tilde{n}_f
\left(\delta\right)\right\}=\frac{1}{2}
\left(\sqrt{1+1/4\omega_{m}^{2}\tau^{2}} -1\right)\,.
\label{eq:Tlimit}%
\end{equation}
The regime $\omega_{m}\tau\ll 1$ is in essence the \textit{adiabatic
  limit} \cite{Law1995}, since the cavity dynamics is much faster than
the mechanical oscillator's period. Several recent experiments fall
into this regime \cite{Kleckner2006,Gigan2006}. Expanding
Eq.~(\ref{eq:Tlimit}) we obtain $\tilde{n}_\textrm{TL}\approx
1/4\omega_m\tau \gg 1$ precluding ground state cooling in this
parameter regime. The corresponding final temperature is of order
$\hbar/k_B\tau$ in complete analogy with the Doppler limit of the
laser cooling of harmonically bound atoms, where the sidebands are not
resolved \cite{Stenholm1986Leibfried2003}.

We turn now to the regime where retardation effects become significant
(\emph{i.e.}~$\omega_{m}\gtrsim 1/\tau$).  This has indeed been
observed in recent experimental work pertaining to both amplification
\cite{Kippenberg2005} and cooling \cite{Schliesser2006,Pinard2006} of
a mechanical oscillator mode. In this regime the optical cavity field
cannot respond instantaneously to the mechanical motion and an
entirely viscous radiation pressure force may arise
\cite{Schliesser2006}. In the frequency domain, this can be
interpreted as having mechanical frequencies which exceed or are equal
to the cavity linewidth [cf.~Fig.~\ref{fig:FP}(c)]. Accordingly, the
asymmetry in the Stokes and anti-Stokes scattering rates becomes more
pronounced leading into the analog of the ``resolved sideband'' limit
of the laser cooling of harmonically bound atoms.  More precisely for
$4\omega_{m}^{2}\tau^{2}\gg1$, Eq.~(\ref{eq:Tlimit}) yields
$\tilde{n}_\textrm{TL}\approx1/16\omega_{m}^{2}\tau^{2}\ll1$ implying
that in this limit one can \emph{reach arbitrarily low temperatures}
(ground state cooling). In a concrete realization, a sound benchmark
to evaluate the cooling performance is whether occupancies below unity
can be attained. Eq.~(\ref{eq:Tlimit}) leads to the following
criterion \footnote{This lower threshold, as compared with atomic
  laser cooling, is related to the absence of a diffusive channel.}:
$\tilde{n}_\textrm{TL}<1\Longleftrightarrow\omega_{m}\tau>1/\sqrt
{32}$; and Eq.~(\ref{eq:ntilde}) implies that within this regime
occupancies below unity are only possible for a certain ``detuning
window'' $-\sqrt{8\omega_{m}^{2}-1/4\tau^{2}}\!\leq\!\Delta_L+
3\omega_{m}\!\leq\!\sqrt{8\omega_{m}^{2}-1/4\tau^{2}}$. 
%(cf.~Fig.~\ref{fig:FT}).    

%with the ratio given by the natural linewidth of the
%mechanical resonance divided by its effective linewidth induced by the
%cooling. 

Finally, we consider the impact of the intrinsic dissipation on the
optimal value of $n_f$ in the regime $\omega_{m}\tau>1/\sqrt{32}$
(analogous considerations can be done for the opposite regime). The
situation is reminiscent of the \textquotedblleft
atomic\textquotedblright\ laser cooling of nanoresonators
\cite{Wilson2004Martin2004,Zoller2004} where the finite $Q_m$ also
plays a crucial role. However, in the present context, the analysis is
simpler given that the optimal laser detuning to maximize the cooling
rate $\Gamma$ [and thus minimize the first term in Eq.~(\ref{eq:nf})]
is of the same order as $\delta_\textrm{opt}$. Hence, the only
relevant issue is the upper bound on $P$ required by the wide
timescale separation underpinning our adiabatic treatment of the
cooling and heating processes. Given the quadratic nature of our
approximate theory, already discussed, the Bose enhancement of the
resonator mode plays no role and the adiabatic requirement reads
$A_{-}\ll1/\tau$.  Aside from the limitations of our approximate
treatment, heuristic considerations imply that the timescale over
which $\langle a_{m}^{\dagger}a_{m}^{\vphantom\dagger}\rangle(t)$
reaches its steady state value can never be shorter than $\tau$. Thus,
our treatment provides an upper bound for the ultimate final
temperature when the finite $Q_{m}$ is considered. As an illustration
we consider the parameters of Ref.~\cite{Schliesser2006}
(\emph{i.e.}~$\Delta_L \tau>0.5$, $\omega_{m}/2\pi=60$~MHz,
$\tau=3$~ns). For a reservoir temperature of $4$~K we have
$n_i\approx1390$. If we consider the improvements in the mechanical
$Q$-values of toroid microcavities due to vacuum operation
($Q_{m}=30,000$) a cooling rate of $2.8$~MHz is then required to reach
$n_i\gamma_m/\Gamma<1$ [cf.~first term in Eq.~(\ref{eq:nf})]
\footnote{We note that $\Gamma\ll 1/\tau$ is satisfied. For
comparison, the maximum cooling rate of Ref.~\cite{Schliesser2006} was
$0.7$ MHz.}.

The cooling process gives rise to photons which have frequencies that
differ from the pump laser ($\omega_L$). Thus it can be studied in an
experiment by measuring the spectrum of the scattered light. As
depicted in Fig.~\ref{fig:FP}(a), we consider a one-sided cavity and
the relevant observable is the output power. The input-output
formalism implies that in the physical representation its spectrum
$S(\omega)$ is given by the Fourier transform of
$e^{i\omega_L\tau}\!\langle[\sqrt{1/\tau_\textrm{ex}}a_{p}^{\dagger}(t+\tau)
+ a_{\textrm{in}}^{\dagger}(t+\tau)] [\sqrt{1/\tau_\textrm{ex}}
a_{p}^{\vphantom\dagger}(t)
+ a_{\textrm{in}}^{\vphantom\dagger}(t)]\rangle_\mathrm{SS}$. In the
shifted representation $a_{p}(t)\!\rightarrow\!  a_{p}(t)+\alpha$ and
the classical input just adds a $c$-number to the cavity steady state
amplitude. Along the lines of our derivation of Eq.~(\ref{eq:master});
$a_{p}^{\vphantom\dagger}(t),\,a_{p}^{\dagger}(t)$ are treated as
environment operators to be reduced to the system operators
$a_{m}^{\vphantom\dagger}(t),\,a_{m}^{\dagger}(t)$ by integrating out
the corresponding Heisenberg equations of motion. A calculation based
on perturbation theory and the theory of quantum Markov processes
\cite{Gardiner2004} then yields \cite{tobepublished}:
\begin{align}\label{eq:spectrum}
  S(\omega)\,\approx\, &
  \frac{\tau}{\tau_\textrm{ex}}\left\{\frac{P}{\hbar\omega_L}
  \left[\frac{\tau_\textrm{ex}}{\tau}-
  \frac{1-\frac{\tau}{\tau_\textrm{ex}}}{\tau^2\Delta_L^2 +
  \frac{1}{4}} \right]\delta\left(\omega-\omega_L\right)
  \right.\nonumber\\ &\left.+\frac{A_-n_f}{\pi}\,
  \frac{\frac{\gamma_\textrm{eff}}{2}}{\left(\omega-\omega_L-\omega_m
  \right)^2 +\frac{\gamma_\textrm{eff}^2}{4}} \right.\\ &\left.+
  \frac{A_+\left(n_f+1\right)}{\pi}\,
  \frac{\frac{\gamma_\textrm{eff}}{2}}{\left(\omega-\omega_L+\omega_m
  \right)^2+ \frac{\gamma_\textrm{eff}^2}{4}}\right\} \,;\nonumber
\end{align}
where the relative order of the corrections is given by $\eta^2$ for
\emph{all} frequencies, and we have normalized to number of photons per unit
time and unit frequency.  Here $A_\mp$ and $n_f$ are given
respectively by Eqs.~(\ref{eq:A}) and (\ref{eq:nf}) and
$\gamma_\textrm{eff}\equiv\gamma_m + \Gamma$. As expected there is
emission of blue-shifted (Anti-Stokes) photons associated to cooling
and red-shifted (Stokes) photons associated to heating. These motional
sidebands have a linewidth determined by the effective damping rate
$\gamma_\textrm{eff}$ and weights ($N_\mp$) determined by the cooling
and heating rates; namely, $N_-=\frac{\tau}{\tau_\textrm{ex}}A_-n_f$
and $N_+=\frac{\tau}{\tau_\textrm{ex}}A_+(n_f+1)$.

The final occupancies can be retrieved by comparing the above spectra
[Eq.~(\ref{eq:spectrum})] for different input powers. The quantity
$[N_+(P)+N_-(P)]P_0/[N_+(P_0)+N_-(P_0)]P$ provides an upper bound for
the ratio $n_f/n_i$. Here we have introduced a ``reference'' low power
$P_0$ for which $\Gamma, A_+\ll\gamma_m$ implying $n_f(P_0)\approx
n_i$, and assumed that the input power $P$ induces appreciable cooling
[i.~e.~$n_f(P)\ll n_i$]. It is important to note that (given $n_i$)
this upper bound provides an \emph{accurate} direct measurement of the
final temperature for $n_f\gg1/2$. On the other hand the worst case
scenario occurs for $\omega_m\tau\gg1$ and $n_f\approx \tilde{n}_f$
where it yields $2n_f$. However for $n_f\lesssim1$ an accurate
measurement is afforded by the quantity
\begin{equation}\label{eq:RP}  
\frac{N_-(P)}{N_+(P)}\frac{N_+(P_0)}{N_-(P_0)}\frac{n_i}{n_i+1}\!\approx\!
  \frac{n_f(P)}{n_f(P)+1}\! =\! \frac{\tilde{C}n_i + P
  A_+}{\tilde{C}\left(n_i+1\right) + P A_-}
\end{equation}
with $\tilde{C}\equiv\hbar\omega_L
\tau_\textrm{ex}\gamma_m(4\tau^2\Delta_L^2 +1)/4\tau^2\eta^2$. Thus,
the high power limit of Eq.~(\ref{eq:RP}) provides a clear signature
for ground state cooling when it can be achieved. This is in stark
contrast to the case of a laser cooled trapped ion where a well
defined thermal reservoir associated to the intrinsic dissipation is
lacking and detailed balance yields $N_-=N_+$. In a realistic scenario
the central peak cannot be regarded as a delta function and the small
relative weight of the sidebands poses an experimental challenge. This
could be overcome by combining state of the art low noise lasers and
high resolution spectroscopy with a suitable lock-in technique. An
advantage, as compared with the trapped ion case, is the larger number
of photons contained by the sidebands for a given $n_f$
\cite{Monroe1995} \footnote{We note that $A_\mp \sim 1$MHz.}.

\begin{comment}
\begin{align}  
  R(P) &
  =\frac{N_-(P)}{N_+(P)}\frac{N_+(P_0)}{N_-(P_0)}\frac{n_i}{n_i+1}\approx
  \frac{n_f(P)}{n_f(P)+1}\nonumber\\& = \frac{\tilde{C}n_i + P
  A_+}{\tilde{C}\left(n_i+1\right) + P A_-},\nonumber
\end{align}
\begin{equation}  
  R(P)=\frac{N_-(P)}{N_+(P)}\frac{N_+(P_0)}{N_-(P_0)}\frac{n_i}{n_i+1}\approx
  \frac{n_f(P)}{n_f(P)+1}.\nonumber
\end{equation}
We note that $n_f(P)/[n_f(P)+1]= (\tilde{C}n_i + P
A_+)/[\tilde{C}\left(n_i+1\right) + P A_-]$
\end{comment}
%\begin{equation} [N_+(P)+N_-(P)]/N_\textrm{back} \sim \eta^2 Q_m
%  (n_f^2/n_i) (\tau/\tau_\textrm{ex})^2\omega_m/\Gamma_B\nonumber
%\end{equation}
%For optimal detuning and in the well resolved sideband limit.

In summary, we have derived a quantum mechanical model of the
temperature limit for cooling using radiation pressure induced
dynamical back-action and shown that ground state cooling can be
achieved as the optical cavity linewidth becomes smaller than the
mechanical frequency, in analogy to atomic sideband cooling. We find
that the threshold to attain occupancies below unity is given by
$\omega_{m}\tau>1/\sqrt{32}$. Furthermore, we have shown how the
spectrum of the optical cavity output could be used to measure the
final temperatures achieved. Our results could apply to other systems
exhibiting dynamical back-action such as an LC circuit with its
capacitance modulated by a mechanical oscillator \cite{Braginsky1977}.

\textsl{Acknowledgements}: TJK acknowledges funding via a Max Planck
Independent Junior Research Group grant, a Marie Curie Reintegration
Grant (IRG), a Marie Curie Excellence Grant (RG-UHQ), and the NIM
Initiative. 

%\bigskip
\bibliographystyle{apsrev}
%\bibliography{RPRef}

%\bigskip

\end{document}